\documentclass[twocolumn,amsmath,amssymb,amsfonts,superscriptaddress,floatfix,showpacs,prl,aps]{revtex4}

\usepackage{graphicx,mathrsfs,flafter}
\usepackage{color}
\usepackage{hyperref}

\begin{document}
\title{Mapping  the phase diagram of strongly interacting matter
}

\author{V. Skokov}
\affiliation{%
GSI Helmholtzzentrum f\"ur Schwerionenforschung, D-64291
Darmstadt, Germany}

\author{K. Morita}
\affiliation{%
GSI Helmholtzzentrum f\"ur Schwerionenforschung, D-64291
Darmstadt, Germany}

\author{B. Friman}
\affiliation{%
GSI Helmholtzzentrum f\"ur Schwerionenforschung, D-64291
Darmstadt, Germany}



\date{\today}

\begin{abstract} 

We employ a conformal mapping to explore the thermodynamics of strongly interacting matter at finite values of the baryon chemical potential $\mu$.  This method allows us to identify the singularity corresponding to the critical point of a second-order phase transition at finite $\mu$, given information only at $\mu=0$. The scheme is potentially useful for computing thermodynamic properties of strongly interacting hot and dense matter in lattice gauge theory. The technique is illustrated by an application to a chiral effective model.
\end{abstract}

\maketitle

\section{Introduction}

The properties of strongly interacting matter 
at finite temperature, $T\neq0$, and baryon density, $n_{B}\neq 0$, have been studied extensively in recent years. The phase diagram has been 
explored experimentally in heavy-ion collisions and examined theoretically in 
models as well as in
first-principle calculations of Quantum Chromodynamics (QCD).
In the nonperturbative regime of QCD, Lattice Gauge Theory (LGT) provides a powerful tool for computing the thermodynamic properties of strongly interacting matter. 

In LGT calculations, the Feynman path integral 
for the partition function of QCD in Euclidean space-time is evaluated using Monte-Carlo sampling. 
The quark degrees of freedom are integrated out, yielding 
an effective action for the gluons, which includes the so-called fermionic 
determinant. For nonzero (real) values of the baryon 
chemical potential $\mu$, the determinant is complex, which implies that the Monte-Carlo weight function is not positive definite and 
consequently that this technique fails for $\text{Re}\,\mu\neq 0$. 
This is the so-called sign problem, which has impeded 
progress in lattice calculations at finite baryon density.  
  
Various indirect approaches have been
developed, to sidestep the sign problem~(for a review see Ref.~\cite{Philipsen:2005mj}). These generally involve an extrapolation from ensembles, where the fermion determinant is positive definite, e.g. from $\mu=0$ or from imaginary $\mu$. Using such techniques, thermodynamic functions can be computed for small real $\mu/T$, i.e.,  for small baryon densities. 

The extrapolation from the $\mu=0$ ensemble can be performed, 
e.g. by means of a Taylor expansion~\cite{Allton:2002zi} in $\mu/T$.
Without refinements, this method is applicable only within the radius
of convergence of the series, $R^{\mu}$. Also for imaginary $\mu$, the fermion determinant is strictly 
positive definite and consequently systematic LGT calculations are possible~\cite{Roberge:1986mm,Alford:1998sd,deForcrand:2002ci,D'Elia:2002gd}. 
An analytic continuation to real $\mu$ by means of a polynomial
is valid only within the convergence radius of the Taylor
expansion~\cite{deForcrand:2002ci}. Thus, both methods are
restricted in applicability by $R^{\mu}$. 

The radius of convergence of the Taylor series is limited by the distance to the closest singularity in the 
complex $\mu$ plane. Conversely, the convergence properties of a power series yields information on the singularities of the original function. Of particular interest are those singularities corresponding 
to the critical point of a second-order phase transition,  
to a crossover transition~\cite{Hemmer52,Stephanov:2006dn} or
to a spinodal line~\cite{Stephanov:2006dn}. We also consider the
``thermal'' singularities associated with zeros of the inverse Fermi-Dirac 
function~\cite{Karbstein:2006er}. 

The convergence properties of the Taylor series in $\mu$ 
has been studied in model calculations~\cite{Wagner:2010,Karbstein:2006er}. 
It was found that on the order of 20 terms or more are needed
to obtain reliable information on the structure of the phase diagram
for this model. This is well beyond what is presently available in LGT
calculations for small quark
masses~\cite{Gavai:2008zr,Schmidt:2010xm}. Here more refined methods
for the analysis of a truncated power series, based e.g. on the
Pad\'{e} conjecture (for an application to QCD in Ref.~\cite{Lombardo:2005ks}) or on the method presented in this
paper, may prove useful.

In this article we present a method for unravelling a singularity in the complex $\mu$ plane, connected with a second-order phase transition, given a finite number of terms in a series expansion of some thermodynamic function. The method, which utilizes a conformal mapping of the Taylor expansion in $\mu$, yields reliable results already at fairly low orders of the series.  Such methods have been employed in quantum field theory for analytic continuation of
perturbative results to the strong coupling regime~\cite{Kazakov:1978ey}, as well as in condensed matter
physics~\cite{danielian-stevens,domb-sykes}, where conformal mapping techniques were used to enhance the sensitivity
to the properties of the critical point~\cite{pearce}.

{
Methods for analyzing a power series of the type employed in this paper, are based on the theorem of Darboux on asymptotic approximations to the late coefficients of a series~\cite{Henrici}. 
These methods can be used if the physical
singularity, e.g. critical point of a second-order phase transition, is dominant~\cite{pearce}. If this is not the case, it may be possible to transform the series by means of a conformal mapping into a form where the critical point is the closest singularity, thus permitting an analysis based on Darboux's theorem. Such a transformation enhances the sensitivity of the series coefficients to the physical singularity and minimizes the influence of other singularities.

In another class of methods, Pad\'{e} approximants are used to improve the convergence properties of the series.  These methods, which are based on the Pad\'{e} conjecture, are, in principle, not restricted to the case where the physical singularity is dominant. Nevertheless, the results of such an analysis are usually more reliable also in the case of one dominant singularity~\cite{pearce}. In this paper, we explore a method based on Darboux's theorem, but stress that methods of the Pad\'{e} class also profit from a judiciously chosen conformal mapping, which enhances the sensitivity to the critical singularity. 
}

The analytic structure in the complex $\mu$ plane was also explored in 
a simple exactly solvable model in Ref.~\cite{Stephanov:2006dn}, where in particular the role of Lee-Yang zeros in finite systems was discussed. 
In this article, we focus on systems in the thermodynamic limit, and defer the discussion of finite-size effects to a subsequent paper.

\section{Conformal mapping}

The thermodynamic functions are analytic in the cut complex $\mu$ plane. The 
cuts are associated with the singularities mentioned in the introduction. First let us assume that the location and type of the singularities of 
a thermodynamic function in the complex $\mu$ plane are known, but only a finite number of terms in the Taylor expansion about $\mu=0$ can be computed. In this case one 
can employ a suitable conformal mapping to improve the convergence properties of the power series and thus obtain reliable results beyond the radius of convergence of the original expansion~\footnote{In scattering theory conformal mapping has been 
used to extend the applicability of low-energy approximations~\cite{Frazer:1961zz,Gasparyan:2010xz}.}.

\begin{figure}[t]
  \includegraphics*[width=7cm]{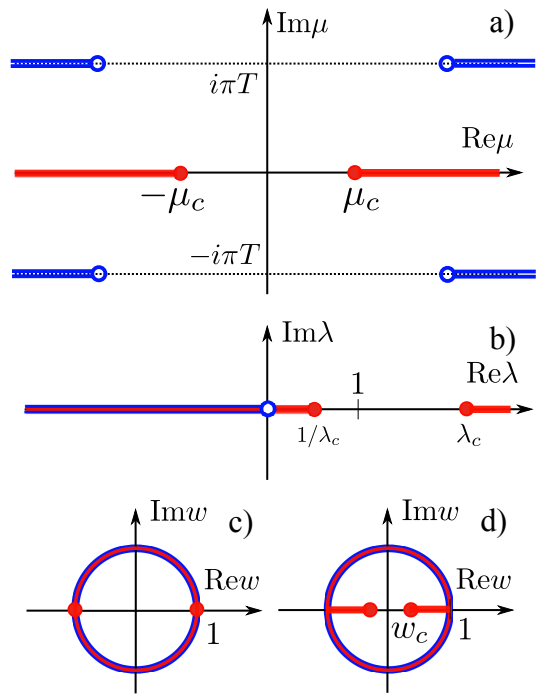}
  \caption {
a) The analytic structure of the order parameter in the complex $\mu$ plane.
The branch points at $\pm \min (\varepsilon_{p}) \pm i \pi T$ are the thermal singularities (open dots), while those 
at $\pm\mu_{\rm c}$ on the real axis correspond to the critical point of the assumed second-order phase transition (filled dots). \newline
b) The analytic structure of the order parameter in the complex fugacity $\lambda$ plane. 
The thermal cut is located between the branch points at $0$ and $-\infty$, 
while the branch points at $\lambda_{\rm c}$ and $1/\lambda_{\rm c}$ are associated with the second-order phase transition. \newline
c) The analytic structure of the order parameter in the complex $w$ plane.  All cuts are  on the circumference
of the unit circle. The branch points corresponding to the critical point are at $w=1$ and $w=-1$, respectively.\newline
d) The analytic structure of the order parameter in the complex $w$ plane for $\lambda_{\rm g}>\lambda_{\rm c}$.  The branch points corresponding to the critical point are located at $w=w_{\rm c}$ and $w=-w_{\rm c}$, respectively.
     }
  \label{fig:ComplexPlane}
\end{figure}

 To illustrate the idea we consider an elementary example, the chiral quark model~\cite{qm} (QM) in the mean-field approximation.
The method is, however, general and applicable to a wide range of problems. 
In the chiral limit, for vanishing chemical potential, $\mu=0$, the system undergoes the second-order phase transition at 
the temperature $T=T_{\rm c}$. In the QM model, the critical temperature
decreases with increasing chemical potential.  
Then, at a temperature $T<T_{\rm c}$ (and larger than  the tricritical temperature), a critical point is located
at some value of the chemical potential $\mu_{\rm c}(T) $ and by symmetry at $-\mu_{\rm c}(T)$ (see Fig. \ref{fig:ComplexPlane}a). Owing to the periodicity of the Fermi-Dirac distribution function
$f_{p}(T,\mu)=1/\{\exp[(\varepsilon_{p}-\mu)/T]+1\}$,
the branch points are 
repeated in the imaginary $\mu$ direction with the period  $2 i \pi T  $. 

In addition to these singularities, there are also thermal branch points, corresponding to the zeros of the inverse Fermi function that are closest to the imaginary axis~\footnote{The gap between the thermal branch point and the real axis is due to the mean-field approximation. In 
general the thermal cuts are images of those of the exact Green function in the complex energy plane, and therefore
 cover the range $-\infty< {\rm Re} \,\mu<\infty$ for ${\rm Im} \,\mu = \pm [\pi T + 2  \pi k T]$ in nonsuperfluid systems.}, $\mu_{\rm th} = \pm [\min(\varepsilon_{p})+i\pi T + 2 i \pi k T],\,k\in \mathcal{N} $. 
In QCD, owing to the $Z(3)$ symmetry (see Ref.~\cite{Roberge:1986mm}) the thermal singularities are
 located at  ${\rm Im}\,\mu = \pm [\pi T + 2  \pi k T]/3 $.  

A more transparent picture emerges when we apply the mapping $\lambda=e^{\mu/T}$, where both the thermal cuts and the copies of the critical branch point are mapped onto the real axis. This transformation amounts to a change of variables from the chemical potential $\mu$ to the fugacity $\lambda$.
In the complex fugacity plane (see Fig.~\ref{fig:ComplexPlane}b) the thermal branch points at  $\mu_{\rm th}$ are mapped onto
the negative real $\lambda$-axis, while  
images of the critical points are located  at $\lambda_{\rm c}=e^{\mu_{\rm c}/T}$ and at $1/\lambda_{\rm c}$. The thermal singularity closest to $\lambda=1$ is the branch point at $\lambda_{\rm th}=0$, corresponding to $\varepsilon_{p}=\infty$. 
 This is a universal  property, valid also beyond the mean-field approximation.
Consequently, since $\lambda_{\rm c}>1$ and hence $0<1/\lambda_{\rm c}<1$,
the singularity closest to $\lambda=1$ 
is the one at  $1/\lambda_{\rm c}$.
Thus, in the complex $\lambda$ plane, the radius of convergence, $R^{\lambda}=1-1/\lambda_{\rm c}$, is not affected 
by the thermal singularities. 

The critical points at $\lambda_{\rm c}$ and 
at $1/\lambda_{\rm c}$ are branch points in the complex $\lambda$ plane.
In the mean-field approximation, the order parameter scales as $\sigma(\mu)\sim (\mu_{\rm c}-\mu)^{1/2}$ for $\mu \lesssim \mu_{\rm c}$. Thus, the analytic continuation of the order parameter exhibits one or more cuts starting at the branch point $\mu=\mu_{\rm c}$. In systems of finite volume, these cuts correspond to lines of singularities, the Lee-Yang zeroes~\cite{Yang:1952be,Hemmer52,Fisher,Grossmann69,Itzykson:1983gb,Stephanov:2006dn}, which generally form a nonzero angle with the real axis. In the thermodynamic limit, we use the freedom to choose the location of the cuts and place them 
on the positive real axis starting at $\lambda_{\rm c}$ and on the real axis in the negative direction starting at $1/\lambda_{\rm c}$, respectively.

We now map the cut fugacity plane onto the interior of the unit circle.
A suitable conformal mapping  is given by 
\begin{equation}
w(\lambda)=\frac{\sqrt{\lambda \lambda_{\rm c}-1}-\sqrt{\lambda_{\rm c}-\lambda}}{\sqrt{\lambda \lambda_{\rm c}-1}+\sqrt{\lambda_{\rm c}-\lambda}}\, .
\label{conf-map}
\end{equation}
The branch points at $\lambda_{\rm c}$ and $1/\lambda_{\rm c}$ are mapped onto $w=1$ and $w=-1$ respectively, while the corresponding cuts are
mapped onto the circumference of the unit circle, as illustrated in Fig.~\ref{fig:ComplexPlane}c. We note that the symmetry $\mu\leftrightarrow -\mu$ corresponds to $w\leftrightarrow -w$ in the mapped variable. A Taylor series about 
$w=0$, which corresponds to $\lambda=1$, converges for all points within the unit circle.  
Consequently, one can use this mapping to construct an analytic continuation of the Taylor series in $\lambda$, which is valid also beyond 
the radius of convergence in the $\lambda$ plane, $R^{\lambda}$.  

We consider first the ideal case, where the dependence of the order parameter on the chemical potential,
or equivalently on the fugacity, is known. Then, using the inverse transformation to (\ref{conf-map}) 
\begin{equation}
\lambda=\frac{\lambda_{\rm c} (1-w)^{2}+(1+w)^{2}}{ (1-w)^{2}+\lambda_{\rm c}(1+w)^{2}}\, ,
\end{equation}
we obtain the order parameter as a function of $w$, 
$\sigma_{w}(w)=\sigma(\lambda(w))$. The corresponding Taylor series in $w$ is given by
\begin{equation}
\sigma_{w}(w)=\sigma_{w}(0)+\sum_{n=1}^{\infty}c^{w}_{n}w^{n}\, ,
\label{Taylor_w}
\end{equation}
where
\begin{equation}
c^{w}_{n}=\left.\frac{1}{n!}\frac{\text{d}^{n}\sigma_{w}}{\text{d} w^{n}}\,\right|_{w=0} .
\end{equation} 
The conformally mapped expansion as a function of $\lambda$, which converges beyond
$R^{\lambda}$, is obtained by introducing the mapping (\ref{conf-map}) in the Taylor expansion~(\ref{Taylor_w})
\begin{equation}\label{conf-map-series}
\sigma_{\rm cm}(\lambda)=\sigma_{w}(w(1))+\sum_{n=1}^{\infty}c^{w}_{n}(w(\lambda))^{n}\, .
\end{equation}
The convergence properties of the expansion (\ref{conf-map-series}) are improved compared to those of the original expansion in $\lambda$ or $\mu$,
because information on the analytical structure of the function is included in the mapping. 

\begin{figure}[t]
  \includegraphics*[width=8cm]{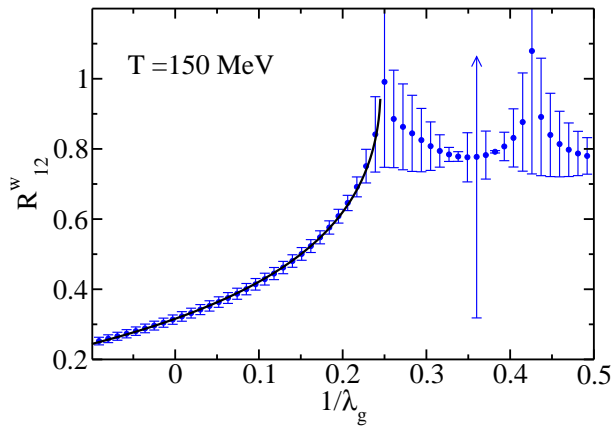}
  \caption { 
The radius of convergence $R_n^w$ in the $w$ plane as a function of $1/\lambda_{\rm g}$ calculated at 
	fixed order $n=12$ (dots). The error bars show the deviation of the radius of convergence  
  $R_{12}^w$ from that obtained with $n=10$,  $R_{10}^w$.   
		The solid line show the result of fitting  with the {\em Ansatz} $R^{w}(\lambda_{\rm g})$.
     }
  \label{fig:Rconv}
\end{figure}

\begin{figure}[t]
  \includegraphics*[width=8cm]{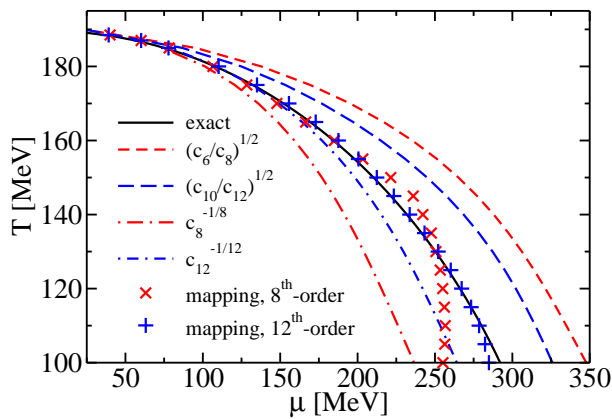}
  \caption {The phase diagram of the QM model in the chiral limit. The  solid line represents 
	the critical line of the second-order phase transition, while the dashed 
	and dash-dotted lines are obtained using different estimates of the radius of convergence based on the original Taylor expansion in $\mu$. The symbols show the phase boundary obtained using the mapping technique discussed the text.
     }
  \label{fig:RestoredPD}
\end{figure}

\section{Results and discussion}

In applications, e.g. to lattice QCD, a strategy by which one can find the location of singularities in the $\mu$ plane, given a finite number of terms in the Taylor expansion, is needed. 
We have found the procedure presented below useful for locating a second-order phase transition, i.e., a critical point on the real $\mu$ axis. In the case of QCD, this could be a critical endpoint, where the crossover transition ends and is replaced by a first-order transition. 

First we define a map of the type (\ref{conf-map}), replacing the critical fugacity $\lambda_{\rm c}$ by a parameter  $\lambda_{\rm g}$
\begin{equation}\label{lambda-g}
w_{\rm g}(\lambda;\lambda_{\rm g})=\frac{\sqrt{\lambda \lambda_{\rm g}-1}-\sqrt{\lambda_{\rm g}-\lambda}}{\sqrt{\lambda \lambda_{\rm g}-1}+\sqrt{\lambda_{\rm g}-\lambda}}\, .
\end{equation}
The analytical structure of the thermodynamic function in the $w_{\rm g}$ plane now depends on the value of $\lambda_{\rm g}$ relative to $\lambda_{\rm c}$. For $\lambda_{\rm g}>\lambda_{\rm c}$, the branch points associated with the critical point, 
$\lambda_{\rm c}$ and $1/\lambda_{\rm c}$, are mapped onto points inside the unit circle at $w=\pm w_{\rm c}=\pm w_{\rm g}(\lambda_{\rm c};\lambda_{\rm g})$ 
(see Fig.\ref{fig:ComplexPlane}\,d).
Since $w=\pm w_{\rm c}$ are the singularities that are closest to the origin, the radius of convergence of the Taylor expansion in $w$ is given by $R^{w}=w_{\rm c}$.
If, on the other hand,
$\lambda_{\rm g}<\lambda_{\rm c}$, the critical point is mapped onto the
circumference of the unit circle and consequently the radius of convergence equals unity, $R^{w}=1$. Thus, given $R^{w}$ as a function of $\lambda_{\rm g}$, 
the location of the critical point, $\lambda_{\rm c}$, can be obtained e.g. by applying the inverse mapping to $w=R^{w}$. Hence, the dependence of the radius of convergence on $\lambda_{\rm g}$ can be used to determine the location 
of a second-order critical point. 

In reality the radius of convergence is known only approximately, because only a finite number of terms of the Taylor expansion are known. In order to enhance the sensitivity to the location of the critical point and minimize the influence of other singularities, it is advantageous to use a mapping which leaves the singularity of interest close to the origin and moves all others as far away as possible~\cite{pearce}. As we show, this can be achieved rather efficiently by varying $\lambda_{\rm g}$ in Eq.~(\ref{lambda-g}). Let us assume that 
the Taylor expansion in $\mu$ or equivalently in $\lambda$ is known up to $n$-th order.  We proceed by 
performing the mapping of the truncated series and obtain an expansion of the type~(\ref{Taylor_w}), truncated to $n$-th order in $w$.
We note that the coefficients $c_{n}^{w}$, are uniquely determined by terms of order $m\leq n$ of the original expansion.

The radius of convergence 
as a function of the parameter $\lambda_{\rm g}$ is approximately given by $R^w_n = |c^{w}_n|^{(-1/n)}$.  (In the limit $n\to\infty$ this expression for the radius of convergence is exact.) 
The approximate expression is, for a given $n$, fitted with the {\em Ansatz} $R^w(\lambda_{\rm g}) = a + w_{\rm g}(\overline{\lambda}_{\rm c};\lambda_{\rm g})$ for $R^{w}<0.95$.
The parameter $a$ accounts for corrections due the truncation at a finite $n$, while $\overline{\lambda}_{\rm c}$ is the approximate critical fugacity. We neglect the dependence of $a$ on $\lambda_{\rm g}$ and justify this approximation {\em a posteriori}.

The resulting fit is illustrated  in Fig.\ref{fig:Rconv} at a temperature $T=150$ MeV for $n=10$ and $12$.
By following the procedure outlined above for temperatures in the range from 100 MeV to $T_{\rm c}$, we obtain an approximative phase diagram of the QM model (see Fig.\ref{fig:RestoredPD}).
For comparison, we also show the results obtained by extracting the radius of convergence directly from the original Taylor series in $\mu$, for $n=8,12$. The present approach yields a 
better estimate for the location of the critical point for all values of $n$ considered.
The fact that over a wide range of temperatures, we find approximately the same phase boundary for 
different $n$ confirms the consistency of the {\em Ansatz} for $R^{w}$.  Thus, at least in the QM model, the $\lambda_{\rm g}$ dependence of $a$ is indeed very weak. The parameter $a$ vanishes in the large $n$ limit as expected, with the leading term proportional to $1/n$.  
 
We have tested this method on trial functions with various types of singularities. In every case it was found to provide an efficient estimate for the location of the singularity. This indicates that our method is applicable to a broad range of problems involving the analysis of truncated power series. 

Another interesting method for locating  a critical point, based on Darboux's theorem for the asymptotic expansion of late Taylor coefficients (see e.g.~\cite{Henrici}), was proposed by Hunter and Guerrieri~\cite{Hunter:1980}. This method has the advantage that it can be used to identify both the critical point and the critical exponent. We find that for a fixed critical exponent, $\beta=1/2$, results of similar quality as those shown in Fig.~\ref{fig:RestoredPD} are obtained, while attempts to determine also the critical exponent were not successful for $n\leq 12$. A detailed study of the two methods will be reported  in a subsequent paper.  

For a physical pion mass, we expect a crossover transition at small values of the chemical potential. 
In this case the critical singularity splits into two conjugate branch points, 
located at complex values of the chemical potential symmetrically with respect to the real axis~\cite{Stephanov:2006dn,Itzykson:1983gb}, $\mu_{\rm co}$ and $\mu_{\rm co}^{\star}$. Therefore, for 
a crossover transition, the radius of convergence of the Taylor expansion is given by the distance of $\mu_{\rm co}$ to the origin, $|\mu_{\rm co}|$.  In the QM model, the resulting radius of convergence decreases continuously, as the temperature is increased from the critical endpoint to the pseudocritical temperature at $\mu=0$ and then increases again. The minimum of the radius of convergence is thus located close to the $\mu=0$ pseudocritical temperature and is not associated with the critical end point, as suggested in Ref.~\cite{Allton:2003vx}.

In this article, we presented an efficient scheme for locating the critical point of a second-order phase transition in the thermodynamical limit. Generalizations of our approach to finite systems and to crossover transitions are under study.

We thank A.~Gasparyan, M. Lutz, K. Redlich, and M. Stephanov for stimulating discussions.
K.~M. and V.~S. acknowledge support of the Frankfurt Institute for Advanced Studies (FIAS).

\end{document}